# MIPS-Core Application Specific Instruction-Set Processor for IDEA Cryptography − Comparison between Single-Cycle and Multi-Cycle Architectures

Ahmad Ahmadi [1], Reza Faghih Mirzaee [2*]

[1] Department of Computer Engineering, West Tehran Branch, Islamic Azad University, Tehran, Iran
[2] Department of Computer Engineering, Shahr-e-Qods Branch, Islamic Azad University, Tehran, Iran
[*] Corresponding Author's E-Mail: r.f.mirzaee@qodsiau.ac.ir

*Abstract:* A single-cycle processor completes the execution of an instruction in only one clock cycle. However, its clock period is usually rather long. On the contrary, although clock frequency is higher in a multi-cycle processor, it takes several clock cycles to finish an instruction. Therefore, their runtime efficiencies depend on which program is executed. This paper presents a new processor for International Data Encryption Algorithm (IDEA) cryptography. The new design is an Application Specific Instruction-set Processor (ASIP) in which both general-purpose and special instructions are supported. It is a single-cycle MIPS-core architecture, whose average Clocks Per Instruction (CPI) is 1. Furthermore, a comparison is provided in this paper to show the differences between the proposed single-cycle processor and another comparable multi-cycle crypto processor. FPGA implementation results show that both architectures have almost the same encoding/decoding throughput. However, the previous processor consumes nearly twice as many resources as the new one does.

# 1. Introduction

General Purpose Processor (GPP) is a programmable device which aims to implement a large number of general applications. It is mainly designed for general purpose computers such as PCs or workstations. It benefits from a good time-to-market and flexibility. However, computation speed, power consumption, and hardware size are the main concerns for GPPs. As a result, they are not suitable for fast real-time applications. On the other hand, a Single Purpose Processor (SPP) is a digital circuit executing a single program. For example, the circuit used in a digital camera is an SPP. It has several design advantages such as fast performance, low power consumption, and small size. Nonetheless, they are not very easy to design and need custom design procedure. Therefore, the design time is also high. Other disadvantages are limited flexibility and difficult reprogramming.

An Application Specific Instruction-Set Processor (ASIP) is a compromise between the flexibility of a GPP and the performance of an SPP [1]. It leads to higher computational efficiencies than a general purpose processor and more flexibility than a fixed-function design. The aim is to tailor the instruction-set to a specific application while it supports the execution of general instructions as well. There are many ASIP architectures dedicated to different applications such as video coding, cryptography, telecommunication, and so forth.

This paper presents a new ASIP crypto processor to be able to perform encryption for the International Data Encryption Algorithm (IDEA) [2]. Encryption is the main constituent in information security, which includes confidentiality, integrity, and availability [3]. IDEA is one of the most successful publicly known algorithms. It has been patented in the United States and in most of the European countries. Despite all the advances in cryptanalysis techniques, IDEA is still considered as one of the strongest cryptography methods. Although there are two major reports of breaking the full 8.5-round IDEA by using *narrow-bicliques* [4] and *meet-in-the-middle* [5], these attacks are not practically feasible due to their huge data and time requirements [6]. Therefore, IDEA is still completely secure for practical usage [6].

Various SPP and ASIC architectures of IDEA are available in the literature, some of which have reached high speed [7, 8]. There are also several papers in this subject comparing different design methodologies. The efficiency of synchronous and asynchronous implementations has been compared in [9]. In another study [10], serial and parallel implementations are compared with each other. IDEA implementation has also been investigated beside other well-known cryptographic algorithms [11]. However, there are few ASIP implementations of IDEA in the literature. One example of an ASIP IDEA crypto processor has been given in [12]. It is based on Von Neumann architecture with a single main memory and shared system bus.

The proposed processor in this paper is based on the MIPS architecture. In contrast with [12], the new design is founded upon Harvard architecture with physically separate storages and signal pathways for instructions and data. It utilizes local system bus to complete the execution of an instruction in only one clock cycle. This paper conducts a detailed comparison between the non-pipelined multi-cycle architecture of [12] and the proposed non-pipelined single-cycle processor. The new design is a low-cost architecture consuming few resources in comparison with its counterpart ASIP processor.

The rest of the paper is organized as follow: In section 2, the IDEA algorithm is briefly reviewed. The new processor is presented in Section 3. The hardware components and units are described in detail. Section 4 includes implementation results and comparisons. Finally, Section 5 concludes the paper.

## 2. IDEA Synopsys

Cryptographic systems are divided into two main categories of symmetric key and public key. In the symmetric cryptosystem, the required encrypting and decrypting subkeys are obtained from a single initial secret key, which is transferred safely before running the cryptosystem. IDEA is a symmetric algorithm with the initial key size of 128 bits. Figure 1 shows how IDEA works and how a 64-bit plaintext is encoded into ciphertext with the same size [3]. It is composed of eight successive identical rounds followed by a conversion as the final half round (8.5 rounds in total), or the ninth round (Fig. 1). The 64-bit plaintext is divided into four 16-bit blocks ($X_0$, $X_1$, $X_2$, and $X_3$). In addition, there are six subkeys within each round ($K_1^i$ to $K_6^i$, $1 \leq i \leq 8$). The last round needs four subkeys as well ($K_1^9$, $K_2^9$, $K_3^9$, and $K_4^9$). Thus, 52 subkeys are totally required to complete the entire algorithm. They are extracted from the initial 128-bit key. Finally, $Y_0$, $Y_1$, $Y_2$, and $Y_3$ are the corresponding ciphertext of $X_0$, $X_1$, $X_2$, and $X_3$ in Fig. 1.

*2.1. IDEA Operations*

Due to the special interleaving operations that are used along with Exclusive-OR (XOR), the algorithm confusion and complexity is high. Thus, a small change in input would lead to strong change in the output (Avalanche Effect). IDEA has the following operations:

- Bitwise XOR denoted by $\oplus$
- Modulo-$2^{16}$ Addition denoted by $\boxplus$
- Modulo-$2^{16}+1$ Multiplication denoted by $\odot$

The above operations function on the unsigned integer set of numbers from 0 to 65535, where the all-zero word (0x0000) is interpreted as $2^{16}$. Performing modular addition is as simple as doing normal addition of two 16-bit input variables without considering final output carry. Therefore, the result remains the same in size (16 bits). Modular multiplication is a more complex operation than modular addition. The final result of modulo-multiplication is the remainder of dividing the multiplication result by $2^{16}+1$.

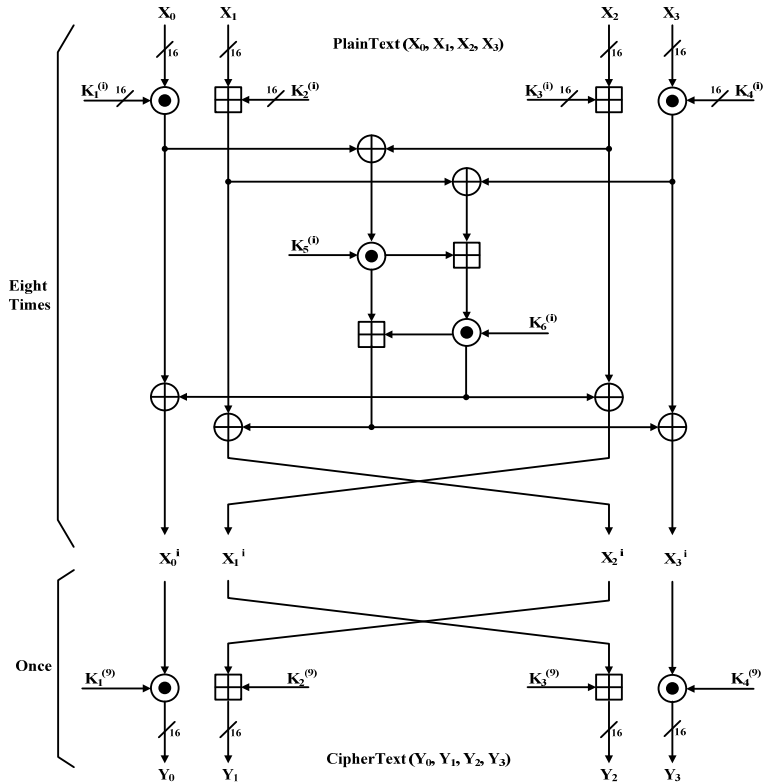

Fig. 1. IDEA algorithm [3]

TABLE I: IDEA Decryption Subkeys

| round i | i = 1 | $2 \leq i \leq 8$ | i = 9 |
|---|---|---|---|
| $K'^i_1$ | $(K_1^{(10-i)})^{-1}$ | $(K_1^{(10-i)})^{-1}$ | $(K_1^{(10-i)})^{-1}$ |
| $K'^i_2$ | $-K_2^{(10-i)}$ | $-K_3^{(10-i)}$ | $-K_2^{(10-i)}$ |
| $K'^i_3$ | $-K_3^{(10-i)}$ | $-K_2^{(10-i)}$ | $-K_3^{(10-i)}$ |
| $K'^i_4$ | $(K_4^{(10-i)})^{-1}$ | $(K_4^{(10-i)})^{-1}$ | $(K_4^{(10-i)})^{-1}$ |
| $K'^i_5$ | $K_5^{(9-i)}$ | $K_5^{(9-i)}$ | - |
| $K'^i_6$ | $K_6^{(9-i)}$ | $K_6^{(9-i)}$ | - |

## 2.2. Key Schedule

As mentioned earlier, 52 subkeys are generated from a 128-bit initial key, which is itself equal to the first eight encrypting subkeys ($K_1^1$ to $K_6^1$, $K_1^2$, and $K_2^2$). The remaining 44 subkeys are generated by rotating the initial key by 25 bits to the left. The only difference between encryption and decryption process is the value of the subkeys. The decrypting subkeys (K's) are obtained from the ones used for encryption, according to Table 1, where $-K_j^i$ and $(K_j^i)^{-1}$ are additive and multiplicative inverse of the subkey $K_j^i$ $1 \leq j \leq 6$, respectively. In mathematics, the additive inverse of a number $x$ is the number that, when added to $x$, yields zero. It can be calculated by Eq. 1 in our subject.

$$-K_i = (2^{16} - K_i) \mod 2^{16} = (2^{16} - K_i) \, \& \, 0xFFFF \tag{1}$$

The multiplicative inverse of a number $x$ is the number that, when multiplied by $x$, yields one. There are several ways to calculate it. It can be extracted by the Extended Euclidean algorithm [13]. Another solution is Binary Extended GCD [3], which does not require division.

## 3. Proposed MIPS-Core Crypto Processor

The proposed processor is shown in Fig. 2. It is based on the MIPS architecture, which is a single-cycle processor. It means that the whole instruction cycle (Fetch-Decode-Execute) completes in only one clock cycle, and the average Clocks Per Instruction (CPI) is 1. The proposed design uses physically separate storage units for instructions and data. It has a 16-bit Arithmetic Logic Shift Unit (ALSU), which is capable of performing some complex operations such as multiplication, division, some modular computations, and different kinds of shift. There is a register back for temporary storage of data. Moreover, a hardwired Control Unit (CU) is implemented through use of combinational logic gates. Finally, all of the components are connected together by a local bus system.

### 3.1. Instruction-Set

ASIP processors support the execution of general-purpose and specific instructions. The new design has two instruction formats (Fig. 3). The R-type instructions (Fig. 3a) are devoted to a set of general logical, arithmetic, shift, and comparison instructions, whose operands are stored in the register bank. The specific instructions for cryptography are also of this type. The I-type instructions (Fig. 3b) include branches, memory access commands, and immediate instructions where one of the operands is indicated within the instruction immediately. The complete instruction-set is available in Appendix.

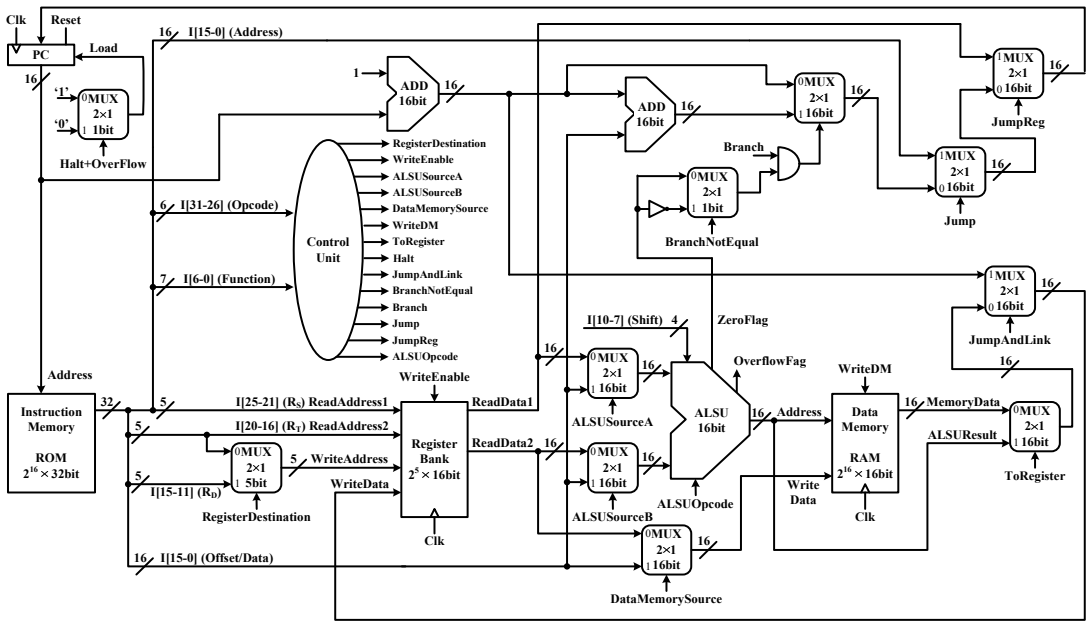

Fig. 2. The proposed MIPS-core crypto processor

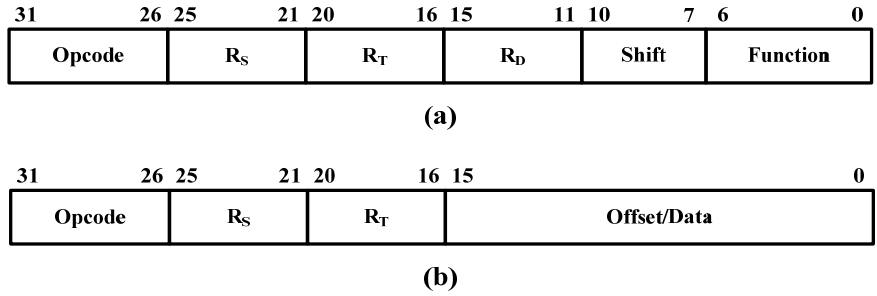

Fig. 3. Instruction formats, (a) R-type, (b) I-type

*3.2. ALSU*

ALSU is the most important part of the processor where different kinds of computations take place. Figure 4 depicts the inside of the proposed ALSU, which is capable of performing signed arithmetic calculations on 16-bit input variables. The list of ALSU operations are depicted in Table 2. The controlling signal *ALSUOpcode* determines what operation must occur.

- *ALSUOpcode = 0 to 3*: Four basic arithmetic operations are considered. Overflow happens if these operations produce a result larger than +32767 or smaller than −32768.

- *ALSUOpcode = 4*: ALSU returns the remainder of dividing the result of an unsigned addition by $2^{16}$. In this case, no overflow occurs and the final result is within the range of 0 to 65535. This operation is supplemented to perform modular addition, which is required in IDEA.

- *ALSUOpcode = 5*: The additive inverse is calculated by Eq. 1. In the topic of IDEA cryptography, the additive inverse is not a negative number, but a positive one whose addition with the input value leads to $2^{16}$, where the least significant 15 bits are zero.

- *ALSUOpcode = 6*: A 16-bit value multiplied by another 16-bit value results in a 32-bit value. However, in a 16-bit arithmetic unit with a signed data range, any result larger than +32767 or smaller than −32768 implies overflow. Therefore, the lower half word (16 bits) is only sent to the output, and the unit must be sensitive to overflow occurrence.

- *ALSUOpcode = 7*: The result is equal to the modulo-$2^{16}+1$ multiplication. There will be no overflow since this operation is dedicated to the IDEA cryptography, not a general arithmetic computation.

- *ALSUOpcode = 8 and 9*: ALSU returns the quotient and remainder parts of an integer division, in which the fractional part is discarded. Integer division is denoted by "A\B" and is equal to $\lfloor A/B \rfloor$.

- *ALSUOpcode = 10 to 15*: Six common shift operations are considered. An equipped barrel shifter is embedded so that ALSU can perform bidirectional arithmetic, logical, and circular shifts. The utilized barrel shifter and rotator is shown in Fig. 5. It is an integrated version of the designs presented in [14, 15]. The core structure shifts/rotates the input number (B) to the left. Data reversal units reverse the bit positions if the input number is supposed to shift/rotate to the right. Besides, in the case of Arithmetic Shift Right (ASHR), sign bit is replicated. In the case of rotation, Most Significant Bits (MSBs) are reentered. Otherwise, zeros are shifted in. Overflow might occur during Arithmetic Shift Left (ASHL). It is the only type of shift which might cause overflow. Finally, there is no output carry in rotation.

- *ALSUOpcode = 16 to 23*: Eight logical operations are performed.

- *ALSUOpcode = 24*: ALSU checks whether the number A is less than B. If so, it returns 0xFFFF (it sets output). If not, it returns 0x0000 (it resets output). To do so, B is subtracted from A, and then, overflow and sign bits are compared. Their inequality implies A<B (Fig. 6).

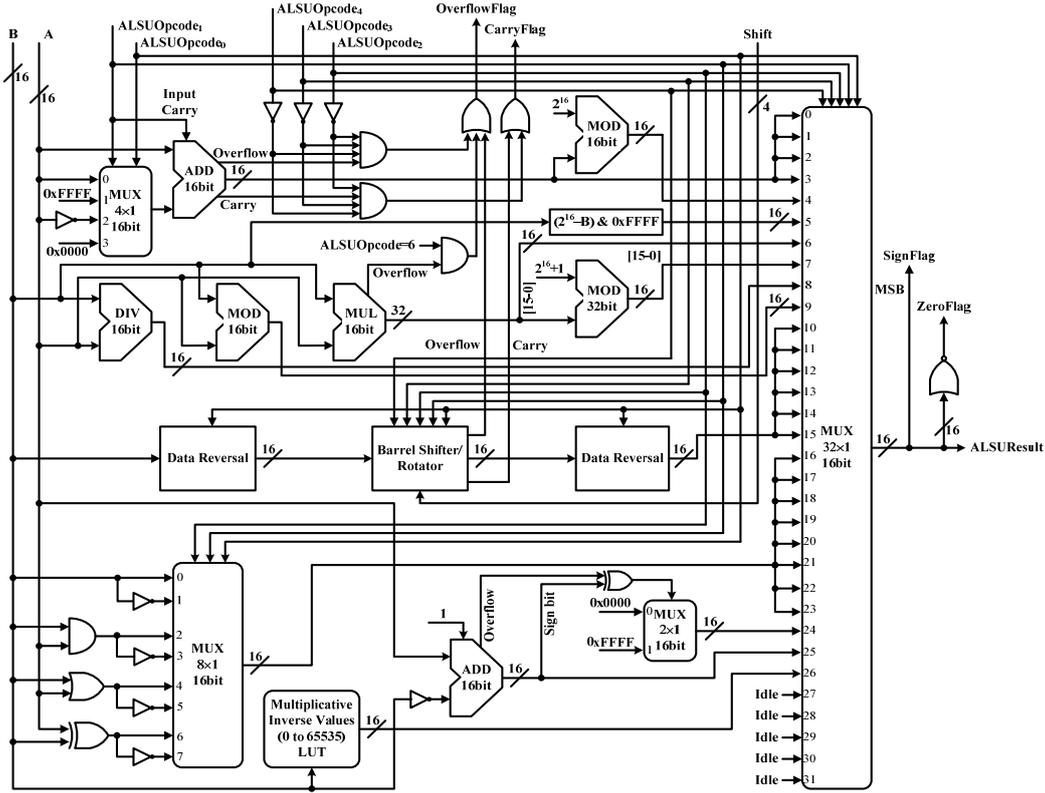

Fig. 4. Inside of ALSU

TABLE II: ALSU Operations

| ALSU Opcode | Unit | Operation | ALSUResult |
|---|---|---|---|
| 0 | | Addition | A+B |
| 1 | | Decrement | A−1 |
| 2 | | Subtraction | A−B |
| 3 | | Increment | A+1 |
| 4 | Arithmetic Unit | Modulo-$2^{16}$ Addition | (A+B) mod $2^{16}$ |
| 5 | | Additive Inverse | ($2^{16}$−B) & 0xFFFF |
| 6 | | Multiplication | A×B |
| 7 | | Modulo-$2^{16}$+1 Multiplication | (A×B) mod $2^{16}$+1 |
| 8 | | Integer Division | A\B |
| 9 | | Remainder | A mod B |
| 10 | | Logical Shift Left | SHL(B) |
| 11 | | Logical Shift Right | SHR(B) |
| 12 | Shift Unit | Rotate Left | ROL(B) |
| 13 | | Rotate Right | ROR(B) |
| 14 | | Arithmetic Shift Left | ASHL(B) |
| 15 | | Arithmetic Shift Right | ASHR(B) |
| 16 | | Transfer | A |
| 17 | | Complement | ¬A |
| 18 | | AND | A∧B |
| 19 | Logic Unit | NAND | ¬(A∧B) |
| 20 | | OR | A∨B |
| 21 | | NOR | ¬(A∨B) |
| 22 | | XOR | A⊕B |
| 23 | | XNOR | A⊙B |
| 24 | | Set on Less Than | if A<B then Set |
| 25 | Arithmetic Unit | Subtraction without Overflow | A−B (without Overflow) |
| 26 | | Multiplicative Inverse | 1/A |

- *ALSUOpcode = 25*: ALSU performs subtraction, but it ignores overflow. This operation is selected whenever conditional branch instructions are executed.

- *ALSUOpcode = 26*: A Lookup Table (LUT) returns the multiplicative inverse for the numbers from 0 to 65535. It takes shorter time to retrieve a value from memory than undergo costly mathematical computations of Binary Extended GCD and Extended Euclidean.

- *ALSUOpcode = 27 and 31*: They are reserved for further development of ALSU in the future.

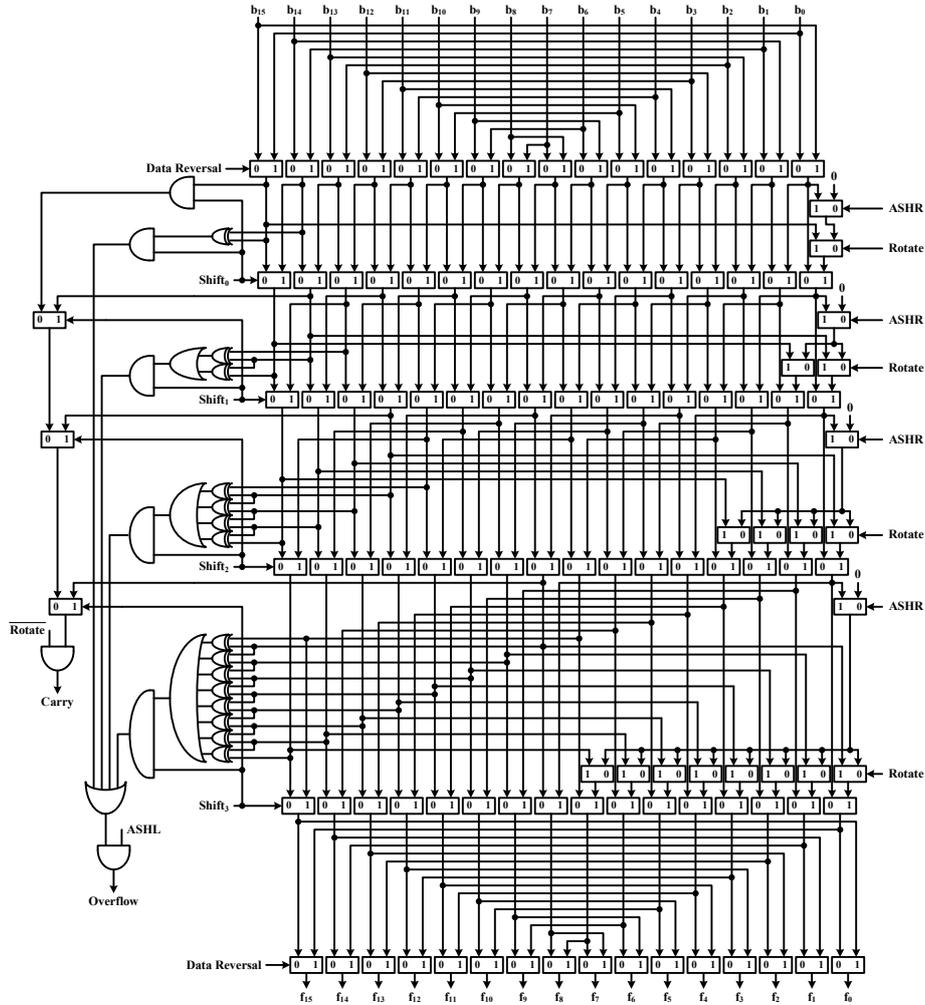

Fig. 5. Utilized bidirectional barrel shifter/rotator

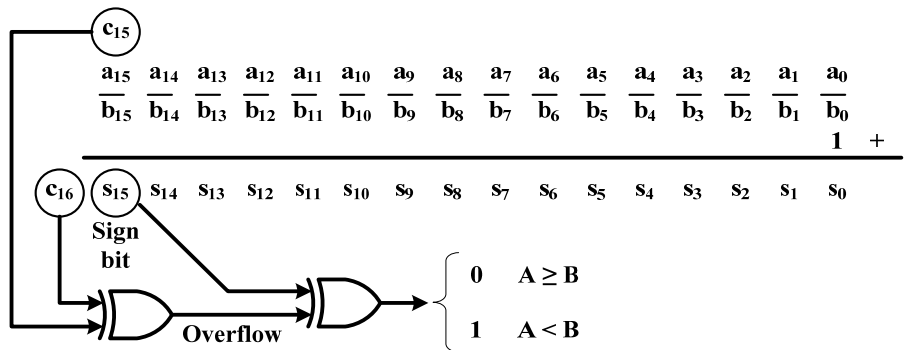

Fig. 6. The way two signed numbers are compared

*3.3. Memory Units*

There are two physically separate memories for instructions and data (Fig. 2). In addition, there is a register bank consisting of 32 registers for temporary storage of data (Fig. 7). Operands are transferred between data memory and register bank via a set of *load* and *store* instructions. It is also worthwhile to mention that all of the storage units including register bank, instruction and data memories, and Program Counter (PC) perform asynchronous memory-read and synchronous memory-write actions. This specification is essentially required to be able to execute an instruction in one clock cycle.

*3.4. Bus System*

Different components in Fig. 2 are connected together by local wires. As a result, data transmission from a unit to another does not block data transmission between other pairs of components. In addition, multiplexers are another important parts of a bus system, which select proper value depending upon which instruction is run.

*3.5. Control Unit*

Hardwired control unit is implemented in this paper through the use of combinational logic gates. For example, the controlling signal *ToRegister* is designed in Fig. 8. It selects between *MemoryData* and *ALSUResult* (Fig. 2). It has to be set to '1' whenever *ALSUResult* is about to be selected. According to the complete instruction-set presented in Appendix, it has to be activated if (*Opcode* = 0 ∧ *Function* = 0 to 24) ∨ (*Opcode* = 1 to 7) ∨ (*Opcode* = 14) ∨ (*Opcode* = 17 to 22).

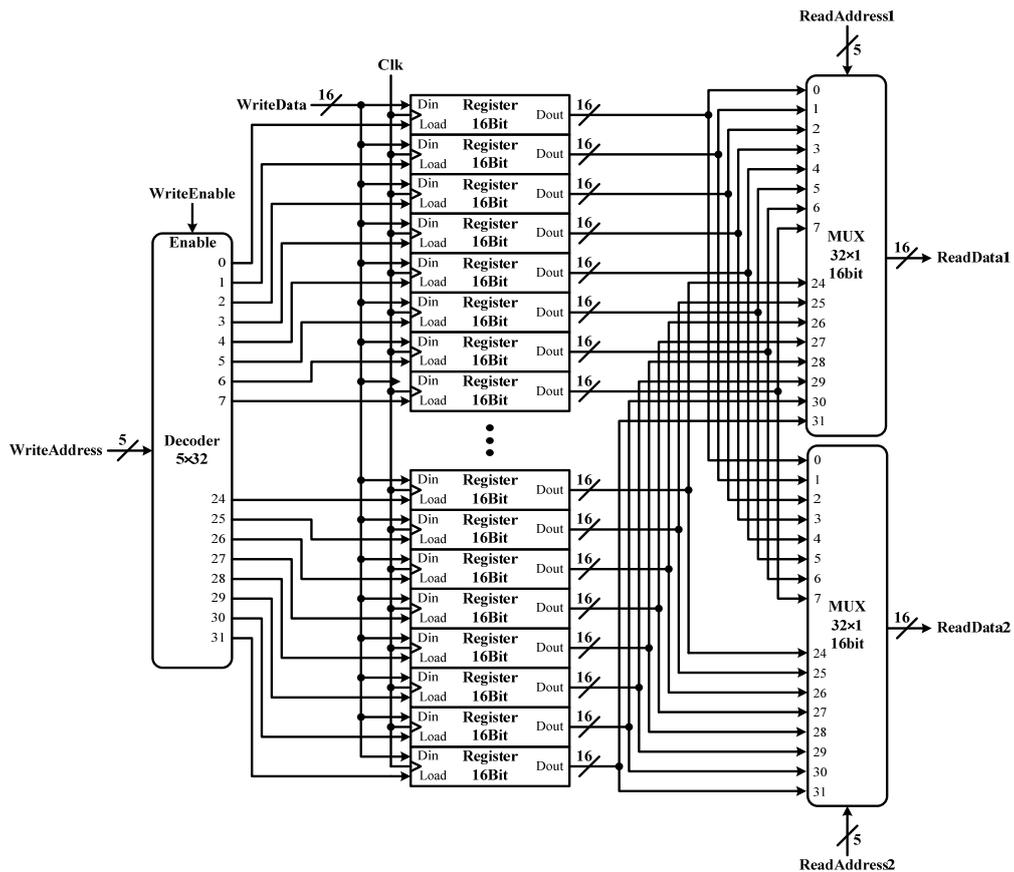

Fig. 7. Inside of register bank

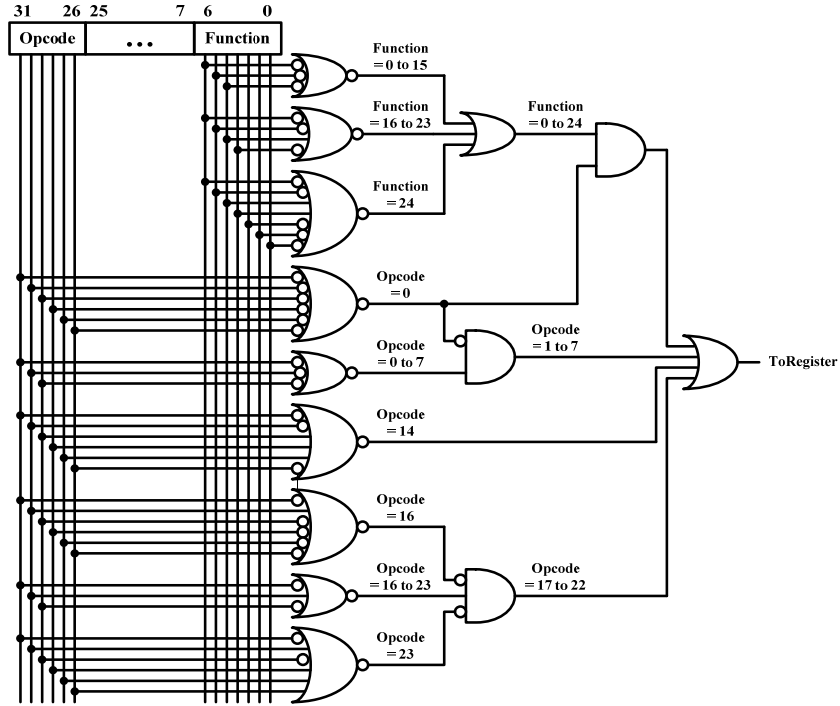

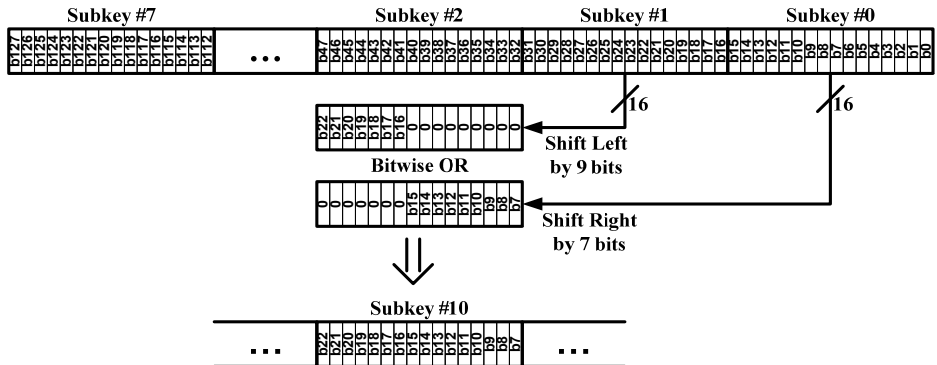

Fig. 8. CU for the generation of *ToRegister*

Fig. 9. Production of subkey #10 from subkeys #0 and #1

*3.6. Execution of Special Instructions for IDEA Cryptography*

Two instructions perform modulo-$2^{16}+1$ multiplication and modulo-$2^{16}$ addition for IDEA cryptography. Another required operation is XOR, which is not a special cryptographic instruction. However, it is frequently used in IDEA rounds (Fig. 1). According to Table 1, additive and multiplicative inverses are also the other special instructions needed for the generation of decrypting subkeys.

At last, rotation is required for the production of encrypting subkeys. As mentioned earlier, the first eight subkeys (Subkey #0 to Subkey #7) are directly extracted from the initial 128-bit key. Then, the initial key is rotated by 25 bits repeatedly until all of the 52 subkeys are generated. Although rotation is supported in the proposed ALSU, the supplemented barrel shifter/rotator (Fig. 5) is only capable of rotating a 16-bit input number by maximum 15 bits. Therefore, it is not feasible to rotate the initial key at once. However, one can use the recursive formula of Eq. 2 to obtain the Subkeys #8 to #51.

The production of the Subkey #10 from Subkeys #0 and #1 is illustrated in Fig. 9 as an example. Subkeys #0 and #1, which are extracted from the initial key, are shifted right by 7 bits and shifted left by 9 bits, respectively. Then, Subkey #10 is equal to their bitwise OR.

$$\text{for } i = 8 \text{ to } 51 \text{ do} \tag{2}$$

$$Subkey\ \#i = ShiftLeft\left(Subkey\ \#\left((i-1)\bmod 8 + \left(\left\lfloor\frac{i}{8}\right\rfloor-1\right)\times 8\right)\right) \vee ShiftRight\left(Subkey\ \#\left((i-2)\bmod 8 + \left(\left\lfloor\frac{i}{8}\right\rfloor-1\right)\times 8\right)\right) \tag{2.1}$$

## 4. Implementation Results and Comparisons

At first, the entire proposed crypto processor is simulated and tested by VHDL code and ModelSim. The encryption and decryption test vectors presented in [3] are applied to test the correct functionality of the processor. The test vectors include a 64-bit plaintext (0000H, 0001H, 0002H, 0003H), a 128-bit initial key (0001H, 0002H, 0003H, 0004H, 0005H, 0006H, 0007H, 0008H), and a 64-bit ciphertext (11FBH, ED2BH, 0198H, 6DE5H). Then, the entire design is synthesized on FPGA platform by Xilinx ISE using Virtex-6 device family. We have also implemented the design in [12] on the same FPGA device. The comparison between the efficiency degree of a single-cycle architecture and a multi-cycle architecture is one of our main targets in this paper.

The specifications of the proposed crypto processor and the one in [12] are exhibited in Table 3. Their implementation results are also shown in Table 4. Although the previous processor has about 71.5% higher clock frequency, the proposed design outperforms its rival in terms of the number of used resources. It uses 46% fewer LUT slices and 56.6% fewer flip flops. Hardware utilization and silicon cost are among the top priorities for ASIP designers [1].

The processor in [12] has shorter clock period (14.795 ns) than the new one (51.911 ns), and hence it seems that it must provide higher speed. However, a multi-cycle design requires several clock cycles to finish an instruction. On the contrary, only one clock cycle is required to complete an instruction in a single-cycle processor. The clock period comes to an end $\frac{51.911}{14.795}$=3.5 times faster in the previous design. However, any instruction with more than 3 clock cycles or any procedure with more than 3.5 times clock cycles will experience longer runtime. For example, four instructions and consequently four clock cycles are required to multiply two numbers in the proposed design. The first two instructions load the operands from data memory to register bank. The third instruction multiplies them, and the last one stores the result into data memory again. The same procedure requires 20 clock cycles in the previous accumulator-based machine. The first instruction, which takes seven clock cycles, loads the accumulator with one of the operands. The second one multiplies the other operand by accumulator. It takes seven clock cycles as well. The last instruction stores the accumulator in six clock cycles. Thus, the whole multiplication procedure takes 4×51.911=207.644 ns and 20×14.795=295.9 ns in the proposed and previous designs, respectively. The time difference is even more significant when performing shift/rotate operations since the previous processor supports only one-bit shift/rotate per instruction.

When it comes to IDEA cryptography, it takes 422 clock cycles to encrypt the first 64-bit plaintext in the proposed design. The procedure includes the generation of subkeys and data encryption. However, the remaining blocks of data can be encrypted in fewer clock cycles (221 clock cycles) because the encrypting subkeys have already been generated. It takes 800 clock cycles to encode the first 64-bit plaintext in the previous crypto processor. The other blocks require 763 clock cycles. Figure 10 shows that both processors perform data encryption with almost the same runtime although the proposed design has lower clock frequency. The time difference becomes even less significant when a large amount of data is encoded. The throughput parameter can be calculated by Eq. 3, where $N$ is the No. of processed bits. The difference between throughputs of the proposed and previous processors is negligible (Table 4).

$$Throughput = \frac{N \times ClockFrequency}{\#ClockCycle} \tag{3}$$

TABLE III: Specifications of Crypto Processors

| Specification | Proposed | [12] |
|---|---|---|
| CPI Type | Single-Cycle | Multi-Cycle |
| Machine Type | Register-Based | Accumulator-Based |
| BUS Type | Local | Shared |
| CU Type | Hardwired | Hardwired |
| Main Memory Type | Separated Instruction and Data Memories | Single Instruction and Data Memory |
| No. of Instruction Formats | 2 | 2 |
| Supports Multiplication | YES | YES |
| Supports Division | YES | NO |
| Supports I/O Instructions | NO | YES |
| Barrel Shifter Inclusion | YES (Multiple-bit Shift/Rotate per Instruction) | NO (One-bit Shift/Rotate per Instruction) |
| Number System | Signed Integer | Unsigned Integer |

TABLE IV: Implementation Results of Crypto Processors

| Parameter | Proposed | [12] |
|---|---|---|
| Clock Period | 51.911 ns | 14.795 ns |
| Maximum Clock Frequency | 19.264 MHz | 67.590 MHz |
| Encoding/Decoding Throughput | 5.578 Mbps | 5.669 Mbps |
| Number of Slice LUTs | 38922 | 72071 |
| Number of LUT Flip Flop Pairs Used | 39450 | 91045 |
| Number of Bounded IOBs | 50 | 52 |

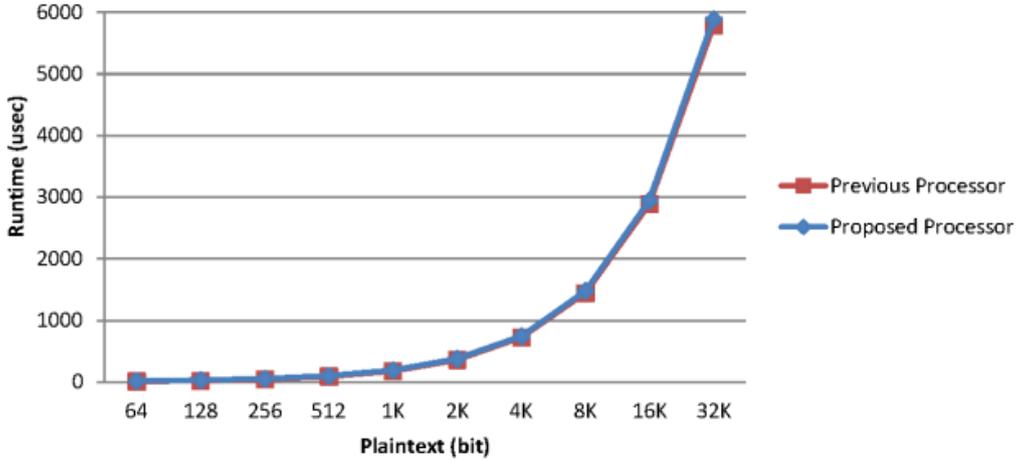

Fig. 10. Encryption runtime versus amount of data

## 5. Conclusion

A new crypto processor has been presented in this paper. It is an ASIP processor whose instruction-set includes both general-purpose and special instructions. The new design completes the execution of every instruction in one clock cycle. Although the clock period is longer, the throughput of the new design is totally comparable to the previously presented multi-cycle processor. The reason is that it completes procedures in fewer clock cycles. In terms of hardware efficiency, the presented processor approximately cuts resource usage by half.

# APPENDIX
## TABLE V: Complete Instruction-Set

| Instruction | Description | Syntax | RTL | Format | Opcode | Function |
|---|---|---|---|---|---|---|
| AND | Bitwise AND | and $RD,$RS,$RT | RD ← RS∧RT | R-Type | 0 | 0 |
| ANDI | AND Immediate | andi $RT,$RS,Data | RT ← RS∧Data | I-Type | 1 | - |
| NAND | Bitwise NAND | nand $RD,$RS,$RT | RD ← RS↑RT | R-Type | 0 | 1 |
| NANDI | NAND Immediate | nandi $RT,$RS,Data | RT ← RS↑Data | I-Type | 2 | - |
| OR | Bitwise OR | or $RD,$RS,$RT | RD ← RS∨RT | R-Type | 0 | 2 |
| ORI | OR Immediate | ori $RT,$RS,Data | RT ← RS∨Data | I-Type | 3 | - |
| NOR | Bitwise NOR | nor $RD,$RS,$RT | RD ← RS↓RT | R-Type | 0 | 3 |
| NORI | NOR Immediate | nori $RT,$RS,Data | RT ← RS↓Data | I-Type | 4 | - |
| XOR | Bitwise XOR | xor $RD,$RS,$RT | RD ← RS⊕RT | R-Type | 0 | 4 |
| XORI | XOR Immediate | xori $RT,$RS,Data | RT ← RS⊕Data | I-Type | 5 | - |
| XNOR | Bitwise XNOR | xnor $RD,$RS,$RT | RD ← RS⊙RT | R-Type | 0 | 5 |
| XNORI | XNOR Immediate | xnori $RT,$RS,Data | RT ← RS⊙Data | I-Type | 6 | - |
| INV | Invert | inv $RD,$RS | RD ← ~RS | R-Type | 0 | 6 |
| INVI | Invert Immediate | invi $RT,Data | RT ← ~Data | I-Type | 7 | - |
| SHL | Shift Left | shl $RD,$RT,Shift | RD ← RT<<Shift | R-Type | 0 | 7 |
| SHR | Shift Right | shr $RD,$RT,Shift | RD ← RT>>Shift | R-Type | 0 | 8 |
| ASHL | Arithmetic SHL | ashl $RD,$RT,Shift | RD ← RT<<<Shift | R-Type | 0 | 9 |
| ASHR | Arithmetic SHR | ashr $RD,$RT,Shift | RD ← RT>>>Shift | R-Type | 0 | 10 |
| ROL | Rotate Left | rol $RD,$RT,Shift | RD ← RT<<∘Shift | R-Type | 0 | 11 |
| ROR | Rotate Right | ror $RD,$RT,Shift | RD ← RT∘>>Shift | R-Type | 0 | 12 |
| BEQ | Branch on Equal | beq $RT,$RS,Offset | If (RS=RT) Then PC ← PC+1+Offset | I-Type | 8 | - |
| BNE | Branch on Not Equal | bne $RT,$RS,Offset | If (RS!=RT) Then PC ← PC+1+Offset | I-Type | 9 | - |
| J | Jump | j Address | PC ← Address | I-Type | 10 | - |
| JR | Jump Register | jr $RS | PC ← RS | I-Type | 11 | - |
| JAL | Jump & Link | jal $RT,Address | RT ← PC+1<br>PC ← Address | I-Type | 12 | - |
| LW | Load Word | lw $RT,Offset($RS) | RT ← M[RS+Offset] | I-Type | 13 | - |
| LWI | LW Immediate | lw $RT,Data | RT ← Data | I-Type | 14 | - |
| SW | Store Word | sw $RT,Offset($RS) | M[RS+Offset] ← RT | I-Type | 15 | - |
| SWI | SW Immediate | sw $RS,Data | M[RS] ← Data | I-Type | 16 | - |
| ADD | Addition | add $RD,$RS,$RT | RD ← RS+RT | R-Type | 0 | 13 |
| ADDI | ADD Immediate | addi $RT,$RS,Data | RT ← RS+Data | I-Type | 17 | - |
| SUB | Subtraction | sub $RD,$RS,$RT | RD ← RS−RT | R-Type | 0 | 14 |
| SUBI | SUB Immediate | subi $RT,$RS,Data | RT ← RS−Data | I-Type | 18 | - |
| MUL | Multiplication | mul $RD,$RS,$RT | RD ← RS×RT | R-Type | 0 | 15 |
| MULI | MUL Immediate | muli $RT,$RS,Data | RT ← RS×Data | I-Type | 19 | - |
| DIV | Division | div $RD,$RS,$RT | RD ← RS\RT | R-Type | 0 | 16 |
| DIVI | DIV Immediate | divi $RT,$RS,Data | RT ← RS\Data | I-Type | 20 | - |
| MOD | Remainder | mod $RD,$RS,$RT | RD ← RS mod RT | R-Type | 0 | 17 |
| MODI | MOD Immediate | modi $RT,$RS,Data | RT ← RS mod Data | I-Type | 21 | - |
| INC | Increment | inc $RD,$RS | RD ← RS+1 | R-Type | 0 | 18 |
| DEC | Decrement | dec $RD,$RS | RD ← RS−1 | R-Type | 0 | 19 |
| SLT | Set on Less Than | slt $RD,$RS,$RT | If (RS<RT) Then RD ← 1 Else RD ← 0 | R-Type | 0 | 20 |
| SLTI | SLT Immediate | slti $RT,$RS,Data | If (RS<Data) Then RT ← 1 Else RT ← 0 | I-Type | 22 | - |
| ADDM | Modular Addition | addm $RD,$RS,$RT | RD ← (RS+RT) mod $2^{16}$ | R-Type | 0 | 21 |
| MULM | Modular Multiplication | mulm $RD,$RS,$RT | RD ← (RS×RT) mod $2^{16}+1$ | R-Type | 0 | 22 |
| ADI | Additive Inverse | adi $RD,$RT | RD ← ($2^{16}$− RT) mod $2^{16}$ | R-Type | 0 | 23 |
| MUI | Multiplicative Inverse | adi $RD,$RS | RD ← 1/RS | R-Type | 0 | 24 |
| HLT | Halt | halt | PC ← Disable | R-Type | 0 | 25 |